\documentclass[10pt,conference]{IEEEtran}
\IEEEoverridecommandlockouts
\usepackage[compress]{cite}
\usepackage[utf8]{inputenc}
\usepackage{amsmath}
\usepackage{amssymb}
\usepackage{mathrsfs}
\usepackage{enumerate}
\usepackage{adjustbox}
\usepackage{xcolor}
\usepackage{graphicx}
\usepackage{algorithm}
\usepackage{algpseudocode}
\usepackage{xpatch}
\usepackage{mathtools}
\usepackage{nicefrac}
\usepackage{float}
\floatstyle{plaintop}
\restylefloat{table}
\usepackage{cleveref}

\algdef{SE}[DOWHILE]{Do}{doWhile}{\algorithmicdo}[1]{\algorithmicwhile\ #1}%
\algrenewcommand\alglinenumber[1]{{\sffamily\footnotesize#1}}
\makeatletter
\xpatchcmd{\algorithmic}{\itemsep\z@}{\itemsep=.25ex plus2pt}{}{}
\makeatother

\DeclareMathOperator{\diag}{diag}
\DeclareMathOperator{\blockdiag}{blkdiag}

\newcommand{\herm}{^\mathsf{H}}

\usepackage{environ}         
\usepackage{etoolbox}        
\usepackage[acronym]{glossaries}
\newacronym{1g}{1G}{first-generation}
\newacronym{4g}{4G}{fourth-generation}
\newacronym{5g}{5G}{fifth-generation}
\newacronym{6g}{6G}{sixth-generation}
\newacronym{mimo}{MIMO}{multiple-input multiple-output}
\newacronym{umi}{UMi}{Urban Micro-cellular}
\newacronym{ris}{RIS}{reconfigurable intelligent surface}
\newacronym{siso}{SISO}{single-input single-output}
\newacronym{mmimo}{mMIMO}{massive multiple-input multiple-output}
\newacronym{cfmmimo}{CF-mMIMO}{cell-free massive multiple-input-multiple-output}
\newacronym{isac}{ISAC}{integrated sensing and communication}
\newacronym{sumimo}{SU-MIMO}{single user MIMO}
\newacronym{mumimo}{MU-MIMO}{multi user MIMO}
\newacronym{embms}{eMBMS}{evolved Multimedia Broadcast and Multicast Service}
\newacronym{sca}{SCA}{successive convex approximation}
\newacronym{sinr}{SINR}{signal-to-interference-plus-noise ratio}
\newacronym{ula}{ULA}{uniform linear array}
\newacronym{uaf}{UatF}{\emph{use-and-then-forget}}
\newacronym{mcs}{MCS}{modulation and coding scheme}
\newacronym{dcc}{DCC}{dynamic cooperation clustering}
\newacronym{mrt}{MRT}{maximum ratio transmission}
\newacronym{ipmmse}{IP-MMSE}{improved partial MMSE}
\newacronym{pmmse}{P-MMSE}{partial-minimum mean square error}
\newacronym{pzf}{P-ZF}{partial ZF}
\newacronym{zf}{ZF}{zero-forcing}
\newacronym{mr}{MR}{maximum ratio}
\newacronym{se}{SE}{spectral efficiency}
\newacronym{ee}{EE}{energy efficiency}
\newacronym{ap}{AP}{access point}
\newacronym{cpu}{CPU}{central processing unit}
\newacronym{uc}{UC}{user centric}
\newacronym{sse}{SumSE}{sum spectral efficiency}
\newacronym{mise}{MinSE}{minimum spectral efficiency}
\newacronym{asd}{ASD}{angular standard deviation}
\newacronym{adr}{ADR}{aggregated data rate}
\newacronym{embb}{eMBB}{enhanced mobile broadband}
\newacronym{mmtc}{mMTC}{massive machine type communications}
\newacronym{urllc}{URLLC}{ultra reliable low latency communications}
\newacronym{csi}{CSI}{channel state information}
\newacronym{pmi}{PMI}{precoding matrix indicator}
\newacronym{ri}{RI}{rank indicator}
\newacronym{csi-rs}{CSI-RS}{CSI-reference signal}
\newacronym{cri}{CRI}{CSI-RS resource indicator}
\newacronym{bs}{BS}{base station}
\newacronym{re}{RE}{resource element}
\newacronym{mmwave}{mmWave}{millimeter-wave}
\newacronym{umwave}{$\mu$mWaves}{micrometer waves}
\newacronym{rnn}{RNN}{recurrent neural network}
\newacronym{cnn}{CNN}{convolutional neural network}
\newacronym{ngmn}{NGMN}{next-generation mobile network}
\newacronym{lte}{LTE}{Long Term Evolution}
\newacronym{lte-a}{LTE-A}{Long Term Evolution Advanced}
\newacronym{5gnr}{5G NR}{5G New Radio}
\newacronym{mm}{MM}{mixed mode}
\newacronym{cdf}{CDF}{cumulative distribution function}
\newacronym{phy}{PHY}{physical}
\newacronym{mac}{MAC}{medium access control}
\newacronym{3gpp}{3GPP}{3rd Generation Partnership Project}
\newacronym{fdd}{FDD}{frequency division duplexing}
\newacronym{tdd}{TDD}{time division duplexing}
\newacronym{ofdm}{OFDM}{orthogonal frequency division multiplexing}
\newacronym{ss}{SS}{synchronization signal} 
\newacronym{pss}{PSS}{primary synchronization signal} 
\newacronym{sss}{SSS}{secondary synchronization signal} 
\newacronym{pbch}{PBCH}{physical broadcast channel} 
\newacronym{dmrs}{DMRS}{demodulation reference signal} 
\newacronym{gnb}{gNB}{next generation nodeB} 
\newacronym{rsrp}{RSRP}{reference signal received power} 
\newacronym{rrm}{RRM}{radio resource management} 
\newacronym{srs}{SRS}{sounding reference signal} 
\newacronym{ran}{RAN}{radio access network} 
\newacronym{nn}{NN}{neural network} 
\newacronym{ue}{UE}{user equipment} 
\newacronym{awgn}{AWGN}{additive white Gaussian noise} 
\newacronym{epa}{EPA}{Extended Pedestrian A model}
\newacronym{eva}{EVA}{Extended Vehicular A model}
\newacronym{etu}{ETU}{Extended Typical Urban model}
\newacronym{tdl}{TDL}{tapped delay line}
\newacronym{cdl}{CDL}{clustered delay line}
\newacronym{uma}{UMa}{urban macro-cell}
\newacronym{isd}{ISD}{inter-site distance}
\newacronym{nlos}{NLoS}{non-line of sight}
\newacronym{los}{LoS}{line of sight}
\newacronym{o2o}{O2O}{outdoor-to-outdoor}
\newacronym{o2i}{O2I}{outdoor-to-indoor}
\newacronym{ul}{UL}{uplink}
\newacronym{dl}{DL}{downlink}
\newacronym{ls}{LS}{least squares}
\newacronym{mmse}{MMSE}{minimum mean square error}
\newacronym{snr}{SNR}{signal-to-noise ratio}
\newacronym{mse}{MSE}{mean square error}
\newacronym{nr}{NR}{New Radio}
\newacronym{prb}{PRB}{physical resource block}
\newacronym{scs}{SCS}{subcarrier spacing}
\newacronym{bler}{BLER}{block error rate}
\newacronym{smmmra}{SMMMRA}{subgroup multicast \gls{mamimo} resource allocation}
\newacronym{mmf}{MMF}{max-min fairness}
\newacronym{smmu}{SMMU}{subgroups of multicast \gls{mamimo} users}
\newacronym{gsmma}{GSMMA}{greedy subgroup multicast \gls{mamimo} algorithm}
\newacronym{apa}{APA}{adaptive power allocation}
\newacronym{ms}{MS}{mobile station}
\newacronym{cb}{CB}{conjugate beamforming}

\usepackage{flushend}
\usepackage{tikz}
\usepackage{changepage}
\usetikzlibrary{fadings}
\tikzfading[name=fade out, 
    inner color=transparent!0,
    outer color=transparent!100]
\usetikzlibrary{patterns}
\usetikzlibrary{shadows.blur}
\usetikzlibrary{shapes}

\ifCLASSOPTIONcompsoc
    \usepackage[caption=false, font=normalsize, labelfont=sf, textfont=sf, subrefformat=parens, labelformat=parens]{subfig}
\else
\usepackage[caption=false, font=footnotesize, subrefformat=parens, labelformat=parens]{subfig}
\fi

\makeatletter
\newcommand\fs@betterruled{%
  \def\@fs@cfont{\bfseries}\let\@fs@capt\floatc@ruled
  \def\@fs@pre{\vspace*{5pt}\hrule height.8pt depth0pt \kern2pt}%
  \def\@fs@post{\kern2pt\hrule\relax}%
  \def\@fs@mid{\kern2pt\hrule\kern2pt}%
  \let\@fs@iftopcapt\iftrue}
\floatstyle{betterruled}
\restylefloat{algorithm}
\makeatother

\begin{document}
\title{RIS-Assisted Cell-Free Massive MIMO: \\ RIS-MS Selection in FR1 and FR3
\thanks{This work has been supported by grants SOFIA-WIND (PID2023-147305OB-C33, funded by MICIU/AEI/10.13039/501100011033 and ERDF, EU), brAIn5G (PID2024-161515OA-I00, funded by MICIU/AEI/10.13039/501100011033 and FEDER), POLIGRAPH (PID2022-136887NB-I00, funded by MCIU/AEI/10.13039/501100011033), and the project AI-XCAST6G (ref. F1263, funded by the IMPULSO program for research at the Rey Juan Carlos University).
}}

\author{\IEEEauthorblockN{Alejandro de la Fuente\IEEEauthorrefmark{1}, Fernando Galindo\IEEEauthorrefmark{1}, Uriel García-Bárbulo\IEEEauthorrefmark{1},\\ Sandra-Noemy Arana-Alegre\IEEEauthorrefmark{1},
Jan García-Morales\IEEEauthorrefmark{1}}
\IEEEauthorblockA{\IEEEauthorrefmark{1}Dept. of Signal Theory and Communications, University Rey Juan Carlos, 28942 Fuenlabrada (Madrid), Spain}
Email: alejandro.fuente@urjc.es}

\maketitle

\begin{abstract}
This paper explores the integration of \glspl{ris} into \gls{cfmmimo} networks operating in FR1 and FR3 frequency bands. We present a comprehensive framework for analyzing \gls{ris}-assisted \gls{cfmmimo} systems under realistic propagation conditions, accounting for frequency-dependent characteristics and \gls{ris} configurations. A novel \gls{ris}-user association algorithm is proposed to optimize phase-shift settings by assigning each \gls{ris} to a single user based on \gls{los} connectivity. The system model incorporates spatially correlated Ricean fading channels and employs scalable \gls{pmmse} combining. The numerical results demonstrate that the proposed \gls{ris}-user selection strategy significantly improves the spectral efficiency compared to random or exhaustive \gls{ris} configurations, particularly when the number of \glspl{ris} is moderate. We also analyze the trade-off between training overhead and performance gains, showing that excessive pilot requirements can offset benefits when \gls{ris} density or element count increases. The results highlight the potential of the FR3 bands for \gls{ris}-assisted \gls{cfmmimo}, provided advanced channel estimation techniques are adopted to mitigate overhead. These findings emphasize the importance of intelligent \gls{ris}-user pairing and scalable estimation methods for future 6G deployments.
\end{abstract}

\begin{IEEEkeywords}
Cell-free massive MIMO, RIS, FR3, RIS-MS selection.
\end{IEEEkeywords}
\glsresetall

\section{Introduction}
\label{sec:introduction}

The evolution toward \gls{6g} wireless networks is driven by the need to support ultra-reliable, low-latency communications, massive connectivity, and high spectral and energy efficiency. Traditional cellular architectures, constrained by inter-cell interference and uneven service quality at cell edges, are increasingly inadequate to meet these demands. In response, \gls{cfmmimo} systems have emerged as a promising solution. These systems eliminate cell boundaries by implementing a large number of distributed \glspl{ap} that cooperatively serve users through centralized processing, thus enhancing coverage uniformity and user fairness~\cite{Interdonato2019,2019Femenias,2024Ngo}.

Parallel to this architectural shift, \glspl{ris} have gained traction as a key enabler of smart radio environments. \glspl{ris} consist of passive, low-cost elements capable of dynamically adjusting the phase of incident electromagnetic waves, thus enhancing signal propagation without requiring additional transmit power or active RF chains. Their ability to manipulate the wireless channel makes them particularly attractive for improving coverage in \gls{nlos} scenarios and mitigating interference \cite{2020Wu,2024Bjornsonbook,2019Renzo,2025delaFuente}.

Integration of \glspl{ris} into \gls{cfmmimo} systems represents a synergistic advancement in wireless communication. \Gls{ris}-assisted \gls{cfmmimo} networks can reshape the propagation environment to support more robust and energy-efficient communication, especially in dense urban or obstructed environments. Recent studies have shown that such integration can significantly improve system performance metrics, including achievable rate, energy efficiency, and secrecy capacity \cite{2019Wu,2021Zhang,2024Zhang}.

The introduction of \glspl{ris} into \gls{cfmmimo} networks poses new challenges in \gls{ris}-\gls{ms} selection, which critically affect overall network performance, such as determining how to associate \glspl{ms} with each \gls{ris} \cite{2024Shi,2023Bie}.

A critical aspect of the deployment of these technologies is selecting the operating frequency bands. The \gls{3gpp} has defined three main frequency ranges for \gls{5g} and beyond: FR1 (sub-$6$ GHz), FR2 ($24$-$52$ GHz, millimeter wave), and the emerging FR3 ($7.125$-$24.25$ GHz, upper midband), which is gaining attention for \gls{6g} research \cite{2021Tataria}. Each band presents unique propagation characteristics and implementation challenges. FR1 offers broad coverage and robust penetration but limited bandwidth, while FR2 enables ultra-high data rates at the cost of higher path loss and limited coverage. FR3, positioned between FR1 and FR2, is considered a \emph{golden band} for \gls{6g}, balancing coverage, capacity, and hardware complexity. Its intermediate wavelength enables dense antenna arrays and more efficient beamforming, making it especially suitable for \gls{cfmmimo} and \gls{ris} deployments in urban and indoor environments \cite{2021Jiang}.

Recent studies highlight that the performance gains of \gls{ris}-assisted \gls{cfmmimo} systems are highly dependent on the propagation characteristics of the chosen frequency band. For example, the effectiveness of \glspl{ris} in manipulating the wireless channel and improving coverage is more pronounced in FR2 and FR3 due to the increased susceptibility of these bands to blockage and multipath effects, which \glspl{ris} can help mitigate \cite{2021Galappaththige}. Moreover, \glspl{ris} in FR3 bands are expected to facilitate advanced applications such as \gls{isac}, further expanding the potential of \gls{cfmmimo} architectures \cite{2024Taghavi}.

{\bf Contributions:} This paper presents a comprehensive framework for the analysis of \gls{ris}-assisted \gls{cfmmimo} systems operating in both FR1 and FR3 frequency bands. The framework enables the evaluation of key performance indicators under realistic deployment conditions, accounting for frequency-dependent propagation and \gls{ris} configurations. A novel \gls{ris}-user association algorithm is proposed to dynamically assign \glspl{ris} to users, thereby improving the system performance in terms of coverage, spectral efficiency, and energy consumption. Furthermore, an extensive set of simulations is conducted to validate the proposed framework and algorithm, providing detailed information on the behavior of \gls{ris}-assisted \gls{cfmmimo} architectures across different frequency bands.

\section{System model}
\label{sec:system_model}

We consider a \gls{cfmmimo} system operating in \gls{tdd} that consists of a \gls{cpu} connected via ideal fronthaul links to $L$ \glspl{ap}, each equipped with $N_\mathrm{L}$ antennas. The \glspl{ap} are deployed over the coverage area to provide data services to $K$ single-antenna \glspl{ms} on the same time-frequency resources \cite{2021Demir}. A set of $R$ \glspl{ris}, operating as antenna arrays equipped with $N_{\mathrm{R}}$ reflective elements (metaatoms), is deployed to increase capacity and provide higher coverage at a low cost, thus improving \gls{cfmmimo} communications. The set of \glspl{ms} is denoted by $\mathcal{K}$ and indexed by $k \!\in\! \mathcal{K}\!=\!\{1,\ldots,K\}$; the set of \glspl{ap} is indicated by $\mathcal{L}$ and indexed by $l \!\in\! \mathcal{L}\!=\!\{1,\ldots,L\}$; and the set of \glspl{ris} is denoted by $\mathcal{R}$ and indexed by $r \!\in\! \mathcal{R}\!=\!\{1,\ldots,R\}$. 
We focus on uplink transmission without loss of generality, as similar principles apply to downlink. Thus, each \gls{tdd} frame is divided into an uplink training phase and an uplink payload data transmission phase, whose lengths, measured in samples, are indicated by $\tau_{\mathrm{p}}$ and $\tau_{\mathrm{u}}$, respectively. The frame length is denoted by $\tau_{\mathrm{c}} \!=\! \tau_{\mathrm{p}} \!+\! \tau_{\mathrm{u}}$ and is assumed to fit the coherence block.

We denote the subset of \glspl{ap} serving \gls{ms} $k$ by $\mathcal{L}_k \subseteq \{1,\ldots, L\}$, where $|\mathcal{L}_k|=L_k \leq L$. For later convenience, given a \gls{ms} $k$, we define the set $\mathcal{S}_k$ as the collection of \glspl{ms} served by some (or all) of the \glspl{ap} serving \gls{ms} $k$, that is, $\mathcal{S}_k = \{i: \mathcal{L}_k \cap\mathcal{L}_i \neq \emptyset\}$. 

\subsection{Channel models}

To evaluate system performance across different frequency bands and consequently, propagation conditions, we consider a spatially correlated Ricean fading channel model with random phase shifts, as proposed in \cite{2019Ozdogan}. This model captures both \gls{los} and \gls{nlos} components, which are essential for FR1 and FR3 bands. A conventional block-fading channel model is considered wherein the channel is time-invariant and frequency-flat within a time-frequency coherence block, and varies independently over different coherence blocks (block fading). The channel vectors $\mathbf{w}_{kl} \in \mathbb{C}^{N_\mathrm{L}}$ (MS–AP), $\mathbf{f}_{kr} \in \mathbb{C}^{N_{\mathrm{R}}}$ (MS–RIS), and matrices $\mathbf{G}_{rl} \in \mathbb{C}^{N_\mathrm{L} \times N_{\mathrm{R}}}$ (RIS–AP) in an arbitrary coherence block can be expressed as
\begin{equation}
\begin{split}
      \mathbf{w}_{kl} \!&=\! \bar{\mathbf{w}}_{kl}e^{j\varphi_{kl}} + \mathbf{w}_{kl}^{\text{NLOS}} ,   \label{eq:w_kl}
\end{split}
\end{equation}  

\begin{equation}
\begin{split}
      \mathbf{f}_{kr} \!&=\! \bar{\mathbf{f}}_{kr}e^{j\varrho_{kr}} + \mathbf{f}_{kr}^{\text{NLOS}} ,   \label{eq:f_kr}
\end{split}
\end{equation}  
and 

\begin{equation}
\begin{split}
      \mathbf{G}_{rl} \!&=\! \bar{\mathbf{G}}_{rl}e^{j\vartheta_{rl}} + \mathbf{G}_{rl}^{\text{NLOS}} ,   \label{eq:G_rl}
\end{split}
\end{equation}  
respectively, where $\bar{\mathbf{w}}_{kl}$, $\bar{\mathbf{f}}_{kr}$, and  $\bar{\mathbf{G}}_{rl}$, denote the deterministic \gls{los} component of the channels, $\varphi_{kl}, \varrho_{kr}, \vartheta_{rl}\sim\mathcal{U}(-\pi,\pi)$ are the random phases of the \gls{los} component of the Ricean channel, and $\mathbf{w}_{kl}^{\text{NLOS}}$, $\mathbf{f}_{kr}^{\text{NLOS}}$, and   $\mathbf{G}_{rl}^{\text{NLOS}}$ represent the \gls{nlos} component. 

Consequently, the aggregated channel  between \gls{ms} $k$ and \gls{ap} $l$ assisted by all \glspl{ris} can be expressed as
\begin{equation}
    \mathbf{h}_{kl} =\underbrace{\mathbf{w}_{kl} \vphantom{\sum_{r=1}^{R} \mathbf{G}_{rl} \mathbf{\Theta}^{\mathsf{H}}_{r} \mathbf{f}_{rk}  }}_{\text{MS-AP}} + \underbrace{\sum_{r=1}^{R} \mathbf{G}_{rl} \mathbf{\Theta}_{r} \mathbf{f}_{kr}  }_{\text{MS-RIS-AP}},
    \label{ec:canal-total}
\end{equation}
where $\mathbf{\Theta}_{r} \in \mathbb{C}^{N_{\mathrm{R}} \times N_{\mathrm{R}}}$ denotes the phase shift matrix in \gls{ris} $r$ as $\mathbf{\Theta}_{r} \triangleq \diag(e^{\theta_{r_1}} \ldots e^{\theta_{r_{N_{\mathrm{R}}}}})$ and $\theta_{r_n} \in [-\pi,\pi]$ represents the phase shift incurred by the metaatom $n$ of \gls{ris} $r$. The first term in (4) represents the direct MS-AP link, while the second term accounts for the cascaded MS-RIS-AP link.

Assuming that the \glspl{ap} and \glspl{ms} are well separated, the channel vectors of different pairs \gls{ms}-\gls{ap}, \gls{ris}-\gls{ap}, and \gls{ms}-\gls{ris} can be considered to be independently distributed, i.e., $\mathbb{E}\{\mathbf{h}_{k'l'} \mathbf{h}_{kl}\herm\} \!=\! \mathbf{0}_{N_\mathrm{L}\times N_\mathrm{L}}, \ \forall k'l' \neq kl$. Thus, the aggregated channel from \gls{ms} $k$ to the complete set of \glspl{ap} $l \in \mathcal{L}$, 
$\mathbf{h}_k\!=\![\mathbf{h}_{k1}^T \ldots \mathbf{h}_{kL}^T]^T$, is distributed~as
$\mathbf{h}_{k}  \sim \mathcal{CN}(\bar{\mathbf{h}}_k,\mathbf{R}_{k})$, where $\bar{\mathbf{h}}_k$ represents the deterministic \gls{los} component of the aggregated channel and $\mathbf{R}_{k} = \blockdiag(\mathbf{R}_{k1},\ldots,\mathbf{R}_{kL}) \in \mathbb{C}^{LN_\mathrm{L}\times LN_\mathrm{L}}$ is the block-diagonal spatial covariance matrix related to \gls{ms} $k$. We define $\mathbf{G}_r\!=\![\mathbf{G}_{r1}^T \ldots \mathbf{G}_{rL}^T]^T$ and $\mathbf{w}_k\!=\![\mathbf{w}_{k1}^T \ldots \mathbf{w}_{kL}^T]^T$ as the channel from \gls{ris} $r$ and the direct channel from \gls{ms} $k$ to the complete set of \glspl{ap} $l \in \mathcal{L}$, respectively.

\subsection{Uplink channel estimation}
\label{sec:channel_estimation}
Configuring a RIS-assisted CF-mMIMO system requires two stages: 
(i) a \emph{separation} channel estimation to acquire the components needed for RIS phase-shift configuration (direct MS-AP vs.\ cascaded MS-RIS-AP), and (ii) an \emph{aggregated} channel estimation that enables combining/precoding in uplink data transmission. We denote by $\mathcal{R}$ the set of RISs and by 
$\mathcal{K}$ the set of MSs. \textcolor{black}{Let $\mathcal{R}_{\mathrm{A}} \subseteq \mathcal{R}$ be the subset of RISs that are assigned to some MS (one MS per RIS) by the selection policy in Sec.~\ref{sec:ris-selection}; by construction, each $r \in \mathcal{R}_{\mathrm{A}}\!=\!\{1,\ldots,R_{\mathrm{A}}\}$ has a LoS link towards its assigned MS. Conversely, RISs with NLoS towards any MS remain unassigned and do not participate in the separation stage.}


\subsubsection{Separation channel estimation}

The aggregated channel can be decomposed into the direct MS-AP link and the cascaded MS-RIS-AP link \cite{2024Shi}. To estimate these contributions for RIS configuration, each RIS is sequentially set into $1+N_{\mathrm{R}}$ distinct states: one for estimating the direct MS-AP link and $N_{\mathrm{R}}$ for the cascaded links through the RIS elements. Whether MSs are processed sequentially or in parallel, the pilot length required to ensure orthogonality must be at least the number of MSs assisted by RIS $r$, denoted by $U_r$. \textcolor{black}{Since each assigned RIS serves exactly one MS, we have $U_r = 1$ for $r \in \mathcal{R}_{\mathrm{A}}$, while $U_r = 0$ for $r \notin \mathcal{R}_{\mathrm{A}}$ (no separation training is performed for unassigned RISs).} Consequently, the separation training overhead (in symbols) satisfies
\begin{equation}
  \tau_{\mathrm{p}_1} \;=\; \sum_{r \in \mathcal{R}} U_r \bigl(1 + N_\mathrm{R}\bigr)
  \;=\; \textcolor{black}{R_{\mathrm{A}} (1 + N_\mathrm{R})}, 
  \label{eq:tau_p1}
\end{equation}
highlighting that resources scale with the number of emph{assigned} RISs and their element count. \textcolor{black}{RISs with NLoS toward any MS are left unassigned and configured with a default low-overhead pattern $\Theta_r = \Phi_r^{(m_r)}$ from a small pseudo-random or orthogonal codebook $\{\Phi_r^{(m)}\}$, without incurring 
any separation pilots.}

\textcolor{black}{An overhead-efficient alternative compatible with our architecture is to perform \emph{subarray-based semi-blind sketching}, where the $N_\mathrm{R}$ RIS elements are partitioned into $N_\mathrm{B}$ uniform blocks and only $1+N_\mathrm{B}$ separation states are required. For example, a $64$-element RIS may be grouped into blocks of $\{4,16,64\}$ elements (i.e., $N_\mathrm{B}=\{16,4,1\}$), which reduces the separation overhead from $1+64=65$ states to $\{17,5,2\}$ states, respectively, while still capturing the dominant structure of the cascaded MS-RIS-AP channel. 
This block-level sketching approach provides a practical semi-blind approximation of the cascaded response and enables a favorable \gls{se}-overhead trade-off without modifying the overall framework.}

\subsubsection{Aggregated channel estimation}

During the uplink training phase, each MS transmits a pilot sequence $\boldsymbol{\psi}_k$ with $\|\boldsymbol{\psi}_k\|^2 = 1$ and mutually orthogonal across MSs. The $N_\mathrm{L} \times \tau_{\mathrm{p}_2}$ received signal at AP $l$ is
\begin{equation}
  \mathbf{Y}_l \;=\; \sqrt{\tau_{\mathrm{p}_2} P_{\mathrm{p}}}\sum_{k \in \mathcal{K}} \mathbf{h}_{kl}\, \boldsymbol{\psi}_k^{\mathsf{T}}
  + \mathbf{N}_l,
  \label{eq:Yl}
\end{equation}
and correlating with $\boldsymbol{\psi}_k^\ast$ yields
\begin{equation}
  \mathbf{y}_{kl} \;=\; \mathbf{Y}_l \boldsymbol{\psi}_k^\ast
  \;=\; \sqrt{\tau_{\mathrm{p}_2} P_{\mathrm{p}}}\,\mathbf{h}_{kl}
  + \sqrt{\tau_{\mathrm{p}_2} P_{\mathrm{p}}}\!\!\!\sum_{i \in \mathcal{P}_k \setminus \{k\}} \!\!\!\mathbf{h}_{il}
  + \mathbf{n}_{kl}.
  \label{eq:yk}
\end{equation}
The \gls{mmse} estimate is
\begin{equation}
  \hat{\mathbf{h}}_{kl} \;=\; \sqrt{\tau_{\mathrm{p}_2} P_{\mathrm{p}}}\,\mathbf{R}_{kl}\, \boldsymbol{\Gamma}_{kl}^{-1}\,\mathbf{y}_{kl},
\end{equation}
where  
\begin{equation}
  \boldsymbol{\Gamma}_{kl} \;=\; \tau_{\mathrm{p}_2} P_{\mathrm{p}} \!\!\sum_{i \in \mathcal{P}_k} \!\!\mathbf{R}_{il} + \sigma_u^2 \mathbf{I}_{N_\mathrm{L}},
  \label{eq:mmse}
\end{equation}
with $\hat{\mathbf{h}}_{kl} \sim \mathcal{CN}(\mathbf{0},\, \tau_{\mathrm{p}_2}P_{\mathrm{p}}\,\mathbf{R}_{kl}\boldsymbol{\Gamma}_{kl}^{-1}\mathbf{R}_{kl})$, 
and the estimation error $\tilde{\mathbf{h}}_{kl} = \mathbf{h}_{kl} - \hat{\mathbf{h}}_{kl}$ is uncorrelated with 
$\hat{\mathbf{h}}_{kl}$. \textcolor{black}{Aggregated-channel estimation uses one orthogonal pilot per MS (i.e., $\tau_{\mathrm{p}_2} = K $), and it is \emph{independent} of whether a given RIS is assigned or not; unassigned RISs do not consume separation pilots (cf.\ 
\eqref{eq:tau_p1}) and keep a fixed low-overhead configuration $\Theta_r = \Phi_r^{(m_r)}$.}

Combining both stages, the total pilot length (in symbols) within the coherence 
block is
\begin{equation}
  \tau_{\mathrm{p}} = \tau_{\mathrm{p}_1} + \tau_{\mathrm{p}_2}
  \;=\; \textcolor{black}{K \;+\ R_{\mathrm{A}} (1 + N_\mathrm{R})}.
  \label{eq:tau_total}
\end{equation}
\textcolor{black}{This formulation also accommodates reduced-dimensional probing, where $N_\mathrm{R}$ is replaced by the number of effective RIS blocks $N_\mathrm{B}$ when block-level grouping is employed. Additionally, (10) makes explicit that only RISs assigned to MSs (and thus having LoS toward them) contribute to the separation overhead. RISs with NLoS to any MS remain unassigned, incur zero separation pilots, and operate with a default configuration, which preserves coherence-block resources while keeping the effective channel stable.}

\subsection{Uplink data combining and spectral efficiency}

During uplink data transmission, the \glspl{ap} in the set $\mathcal{L}$ receive signals from the \glspl{ms} in $\mathcal{K}$ using linear combining techniques. The received signal by \gls{ms} $k$ is expressed as
\begin{equation}
      y_{k} \!=\!  \sum_{l\in\mathcal{L}}\mathbf{h}\herm_{kl}  \mathbf{D}_{kl} \mathbf{v}_{kl}\varsigma_k \!+\! \sum_{l\in\mathcal{L}}\sum^K_{\substack{i=1 \\ i\neq k}} \mathbf{h}\herm_{kl}  \mathbf{D}_{il}\mathbf{v}_{il}\varsigma_i \!+\! n_k,   \label{eq:y_gk}
\end{equation} 
where $n_k \sim \mathcal{CN}(0, \sigma_d^2)$ denotes the receiver \gls{awgn} in \gls{ms} $k$, $\mathbf{v}_{kl} \in \mathbb{C}^{N_\mathrm{L}}$ is the combining vector used by \gls{ap} $l$ for user $k$, and $\varsigma_k$ is the data symbol intended for user $k$, satisfying $\mathbb{E}\{|\varsigma_k|^2\} = 1$ and $\mathbb{E}\{\varsigma_k \varsigma_i^*\} = 0$ for $k \neq i$. The set of diagonal matrices  $\mathbf{D}_{kl} \in \mathbb{C}^{N_\mathrm{L}\times N_\mathrm{L}}$ are used to describe which \glspl{ap} communicate with which \glspl{ms} \cite{2017Buzzi}, and are given by $\mathbf{D}_{kl} \!=\! \mathbf{I}_{N_\mathrm{L}}$ if $l \in \mathcal{L}_{k}$, or $\mathbf{D}_{kl} \!=\!\mathbf{0}_{N_\mathrm{L}\times N_\mathrm{L}}$, otherwise.
The first term in (11) represents the desired signal, the second term accounts for multi-user interference, and the third term is the \gls{awgn}.

In centralized uplink operation, \gls{cpu} has access to the channel estimates of all users, allowing for the joint design of combining vectors.
Let $\hat{\mathbf{h}}_k =\![\hat{\mathbf{h}}_{k1}^T \ldots \hat{\mathbf{h}}_{kL}^T]^T \in \mathbb{C}^{LN_\mathrm{L}}$ be the aggregated channel estimate for user $k$, and
$\mathbf{v}_k \!=\![\mathbf{v}_{k1}^T \ldots \mathbf{v}_{kL}^T]^T \in \mathbb{C}^{LN_\mathrm{L}}$ the corresponding combining vector. The received signal can then be compactly written as
\begin{equation}
      y_{k} \!=\!  \mathbf{h}\herm_{k} \mathbf{D}_k \mathbf{v}_{k}\varsigma_k \!+\! \sum^K_{\substack{i=1 \\ i\neq k}} \mathbf{h}\herm_{k}  \mathbf{D}_i \mathbf{v}_{i}\varsigma_i \!+\! n_k,   
      \label{eq:yk_data}
\end{equation}
where $\mathbf{D}_k\!=\! \blockdiag(\mathbf{D}_{k1},\ldots,\mathbf{D}_{kL}) \in \mathbb{C}^{LN_\mathrm{L}\times LN_\mathrm{L}}$

This work employs the \gls{pmmse} combining, a scalable alternative to the optimal \gls{mmse} combiner, defined as  
\begin{equation}
           \bar{\mathbf{v}}^{\text{P-MMSE}}_k \!=\! {p_k}\Bigg(\sum_{i\in\mathcal{S}_k} p_i \mathbf{D}_k\hat{\mathbf{h}}_{i} (\hat{\mathbf{{h}}}_{i})\herm \mathbf{D}_k+  \mathbf{Z}_{\mathcal{S}_k} + \sigma_{\mathrm{u}}^2 \mathbf{I}_{LN_\mathrm{L}} \Bigg)^{\!\!\!-1}\! \mathbf{D}_k\hat{\mathbf{{h}}}_k,
           \label{eq:pmmse}
\end{equation}
where $p_k$ represents the uplink power for user $k$, $\mathbf{Z}_{\mathcal{S}_k}=\sum_{i\in\mathcal{S}_k}p_i \mathbf{D}_k\mathbf{C}_i\mathbf{D}_k$, and $\mathbf{C}_i$ denoting the error covariance matrix of $\hat{\mathbf{h}}_i$. Unlike centralized \gls{mmse}, \gls{pmmse} relies only on local user sets $\mathcal{S}_k$, reducing fronthaul signaling and complexity.

\textcolor{black}{\textbf{Remark (Unassigned RISs and overhead neutrality).} The aggregated channel vectors $\mathbf{h}_{kl}$ and their MMSE estimates $\hat{\mathbf{h}}_{kl}$ already incorporate the current per-RIS phase state. RISs assigned to some MS employ the per-MS configuration based on the separation stage, whereas RISs \emph{without} LoS to any MS remain unassigned and keep a default low-overhead configuration $\Theta_r=\Phi_r^{(m_r)}$ (cf.\ Secs.~\ref{subsec:ris-selection} and \ref{sec:channel_estimation}). Crucially, these unassigned RISs \emph{do not} consume separation pilots (no contribution to $\tau_{\mathrm{p}_1}$), while their passive state is treated as part of the effective channel realization with no extra signaling burden. Hence, the \gls{pmmse} structure in \eqref{eq:pmmse} remains unchanged, and the \gls{se} analysis below is valid irrespective of whether a given RIS is assigned or operates under the default pattern.}

Combining \eqref{eq:yk_data}–\eqref{eq:pmmse}, the achievable \gls{se} for MS $k$ under standard use-and-then-forget bounding can be expressed in closed form in terms of the channel second-order statistics and the chosen combining rule. (Details are omitted for brevity and follow standard CF-mMIMO derivations.)

\section{RIS-MS selection proposal and RIS phase-shift configuration}
\label{sec:ris-selection}

The integration of \glspl{ris} into \gls{cfmmimo} systems aims to enhance coverage and spectral efficiency, particularly in scenarios with mixed \gls{los} and \gls{nlos} conditions. \Glspl{ris} act as passive arrays that reflect signals toward intended users by adjusting their phase-shift coefficients. However, configuring \glspl{ris} efficiently requires two key steps: (i) selecting which user (or users) each \gls{ris} should assist, and (ii) optimizing the phase-shift matrix for the assigned user (or users). This section introduces a user-centric \gls{ris} assignment strategy and a sequential phase-shift optimization algorithm.

\subsection{RIS-MS selection proposal}
\label{subsec:ris-selection}

The objective of the RIS-MS assignment is to exploit RISs only in operating regimes where they can provide a meaningful deterministic enhancement of the equivalent MS-AP channel, while keeping the associated training overhead manageable. As discussed in Sec.~II-B, the separation training required to configure a RIS for a specific MS scales as $\tau_{\mathrm{p}_1}=U_r(1+N_R)$, which grows linearly with both the number of assigned users and the number of RIS elements. \textcolor{black}{Consequently, dedicating each RIS to a single MS is a deliberate overhead-aware design choice that preserves coherence-block resources and enables scalable multi-RIS operation.}

The assignment begins by ranking users according to the number of APs having a \gls{los} link towards them. Since \gls{los} presence strongly correlates with larger large-scale fading coefficients, users with fewer \gls{los} AP links naturally correspond to weaker direct MS-AP channels. Improving the channel quality of these ``weak'' users yields the largest marginal gain in system coverage and low-percentile spectral efficiency. Hence, these users are considered first.

For each user in this sorted list, the algorithm assigns all RISs that (i) are still unassigned and (ii) have a \gls{los} link towards that user. The rationale is that a RIS can provide a strong deterministic boost to the cascaded MS-RIS-AP path only when a \gls{los} MS-RIS component exists. In the absence of such \gls{los}, the impinging energy at the RIS is dominated by diffuse scattering, which limits the effectiveness of any phase-shift configuration and yields a marginal contribution to the equivalent channel. \textcolor{black}{Therefore, if a given RIS does not have \gls{los} to any MS in the network, it remains unassigned and no small-scale separation training is spent on it. In that case, the RIS applies a default low-overhead configuration $\Theta_r=\Phi_r^{(m_r)}$, selected from a small pseudo-random or orthogonal codebook that does not require channel estimation; the pattern can be kept fixed or slowly hopped over coherence blocks to avoid persistent spatial nulls. Whenever a MS with \gls{los} towards RIS $r$ becomes available, the RIS is immediately eligible for standard per-MS configuration.}

\textcolor{black}{In this selection structure, any RIS that exhibits LoS connectivity to at least one MS is assigned during the main loop, so no LoS-capable RIS remains idle at later stages. Only RISs with NLoS to any MS reach the final stage, where they remain unassigned and operate under the default low-overhead pattern.}

\textcolor{black}{Overall, this procedure directs the RIS resources to the MSs that benefit the most from their presence, while ensuring that training is invested only when a RIS is physically capable of strengthening a MS’s effective channel.} A formal description of the selection policy is provided in Algorithm~\ref{alg:ris-selection}.

\begin{algorithm}[t]
\caption{RIS-MS Selection Based on LoS Connectivity}
\label{alg:ris-selection}
\begin{algorithmic}[1]

\State \textbf{Initialization:}
For each MS $k$, compute the number of APs with LoS towards $k$ and sort MSs in ascending order.

\For{each MS $k$ in the sorted list}
    \For{each RIS $r$ still unassigned}
        \If{LoS$(r,k)$ is true}
            \State Assign RIS $r$ to MS $k$
        \EndIf
    \EndFor
\EndFor
\Statex \textcolor{black}{\textbf{// Policy for RISs with NLoS to any MS}}
\For{each RIS $r$ still unassigned}
    \If{\textbf{no} MS $k$ satisfies LoS$(r,k)$}
        \State \textcolor{black}{\textbf{Do not assign RIS $r$ to any MS}}
        \State \textcolor{black}{$\Theta_r \leftarrow \Phi_r^{(m_r)}$  \ // default low-overhead config}
        \State \textcolor{black}{\textbf{Skip} cascaded channel estimation for RIS $r$}
    \EndIf
\EndFor
\State \textbf{Output:} 
\Statex \hspace{4mm} 1) RIS assignment vector for RISs with LoS-aware pairing  
\Statex \hspace{4mm} \textcolor{black}{2) Default-configuration set $\{\Theta_r\}$ for unassigned RISs}

\end{algorithmic}
\end{algorithm}

\subsection{RIS phase-shift configuration}
\label{subsec:ris-phase-shift}

Once one MS is selected per RIS, the phase-shift matrices of the assigned RISs are optimized to maximize the equivalent MS-RIS-AP channel gain for that MS.
The optimization is performed through a sequential update procedure that adjusts the phase of each reflective element individually. Based on random initial phases uniformly distributed in $[-\pi,\pi]$, the $n$-th element of RIS $r$ is updated as
\begin{equation}
\theta_{r,n} \leftarrow -\operatorname{arg}\big(\mathbf{b}_{r,n}^{\mathsf{H}} \mathbf{A}_{r,n}^{-1} \mathbf{g}_{r,n}\big),
\end{equation}
where $\mathbf{g}_{r,n}$ denotes the channel vector between the $n$-th element of RIS $r$ and the full set of AP antennas, and $\mathbf{b}_{r,n}$ and $\mathbf{A}_{r,n}$ are given in (15)–(16) by
\begin{align}
\mathbf{b}_{r,n} &= p\big(\mathbf{w}_k + \sum_{m\neq n}\mathbf{g}_{r,m}e^{j\theta_{r_m}}f_{kr,m}\big) f_{kr,n}^*,\\
\mathbf{A}_{r_n} &= \mathbf{I}_M + p\big(\mathbf{w}_k + \sum_{m\neq n}\mathbf{g}_{r,m}e^{j\theta_{r,m}}f_{kr,m}\big) \nonumber\\
&\quad \Big(\mathbf{w}_k + \sum_{m\neq n}\mathbf{g}_{r,m}e^{j\theta_{r,m}}f_{kr,m}\Big)^{\mathsf{H}} + p\mathbf{g}_{r,n}f_{kr,n}f_{kr,n}^*\mathbf{g}_{r,n}^{\mathsf{H}},
\end{align}
where $f_{kr_n}$ denotes the channel between \gls{ms} $k$ and the metaatom $n$ of the \gls{ris} $r$.
This iterative process aligns the reflected signals toward the intended MS and reinforces the cascaded MS-RIS-AP link\footnote{Note that this work assumes a perfect separation channel estimation to configure \glspl{ris} phase-shifts, using $\tau_{\mathrm{p}_1}$ channel symbols from the coherence block.} \cite{2024Bjornsonbook}.

\textcolor{black}{
If a RIS has NLoS link towards any MS, it is not assigned to any MS during the selection stage and therefore does \emph{not} undergo the separation channel estimation described in Sec.~II-B. In this case, configuring the phase shifts to maximize a cascaded path that is dominated by NLoS components would provide only marginal benefit while incurring unnecessary training overhead. For such RISs, a default low-overhead configuration is applied as
\begin{equation}
    \Theta_r = \Phi_r^{(m_r)},
\end{equation}
where $\{\Phi_r^{(m)}\}$ is a small pseudo-random or orthogonal codebook that does not require channel estimation. The pattern $\Phi_r^{(m_r)}$ can be kept static or slowly hopped over coherence blocks to avoid persistent nulls. When a MS with a LoS link to RIS $r$ appears, the RIS becomes eligible for the standard per-MS configuration procedure described above.}

This phase-shift policy ensures that each RIS is configured only in scenarios where it can effectively contribute to strengthening the channel of a specific MS, while preserving coherence-block resources by avoiding unnecessary training for RISs that lack LoS connectivity to all MSs.

\textcolor{black}{The proposed RIS-MS selection and configuration scheme is inherently scalable, as it avoids any form of joint active-passive optimization across the entire CF-mMIMO network. Each RIS is dedicated to a single MS, and its phase-shift configuration depends only on the LoS MS-RIS link and on the large-scale statistics of the corresponding MS-AP channels. As a result, RISs operate independently of one another and require no multi-RIS coordination. The CPU only collects lightweight large-scale information-LoS flags, MS-RIS distances, and average pathloss, while the APs estimate the aggregated 
channels locally during the standard uplink training. P-MMSE combining further enhances scalability by relying exclusively on the local user set $\mathcal{S}_k$, without requiring global \gls{csi}. Consequently, the fronthaul and signaling load remain comparable to conventional CF-mMIMO architectures, and 
the proposed approach integrates RISs without creating additional large-scale coordination overhead.}

\section{Numerical results}
\label{sec:results}

We consider a CF-mMIMO network with $L=20$ APs, each equipped with $N_\mathrm{L}=4$ antennas,
uniformly deployed in a $1000 \times 1000$ m$^2$ wrap-around area. The operating bandwidth is $20$ MHz and the channels follow the 3GPP UMi pathloss model with correlated Ricean fading. The coherence block is $\tau_{\mathrm{c}} = 20000$ symbols, \textcolor{black}{providing a quasi-static channel that enables a fair comparison across schemes with different pilot requirements,} and the pilot and data transmit power is $P_{\mathrm{p}} = p_k = 100$ mW. All results are averaged over $100$ random deployments, each with $100$ small-scale realizations.

As described in Sec.~\ref{subsec:ris-selection}, users are sorted according to the number of APs having LoS towards them, and RISs are assigned only when a LoS MS$\leftrightarrow$RIS link exists. \textcolor{black}{Unassigned RISs (those with NLoS to any MS) simply employ the default low-overhead pattern introduced in Sec.~II, and thus do not contribute to the separation overhead.}

\subsection{Sum spectral efficiency vs.\ number of RISs}

\begin{figure}[!t]
\centering
\includegraphics[width=0.5\textwidth]{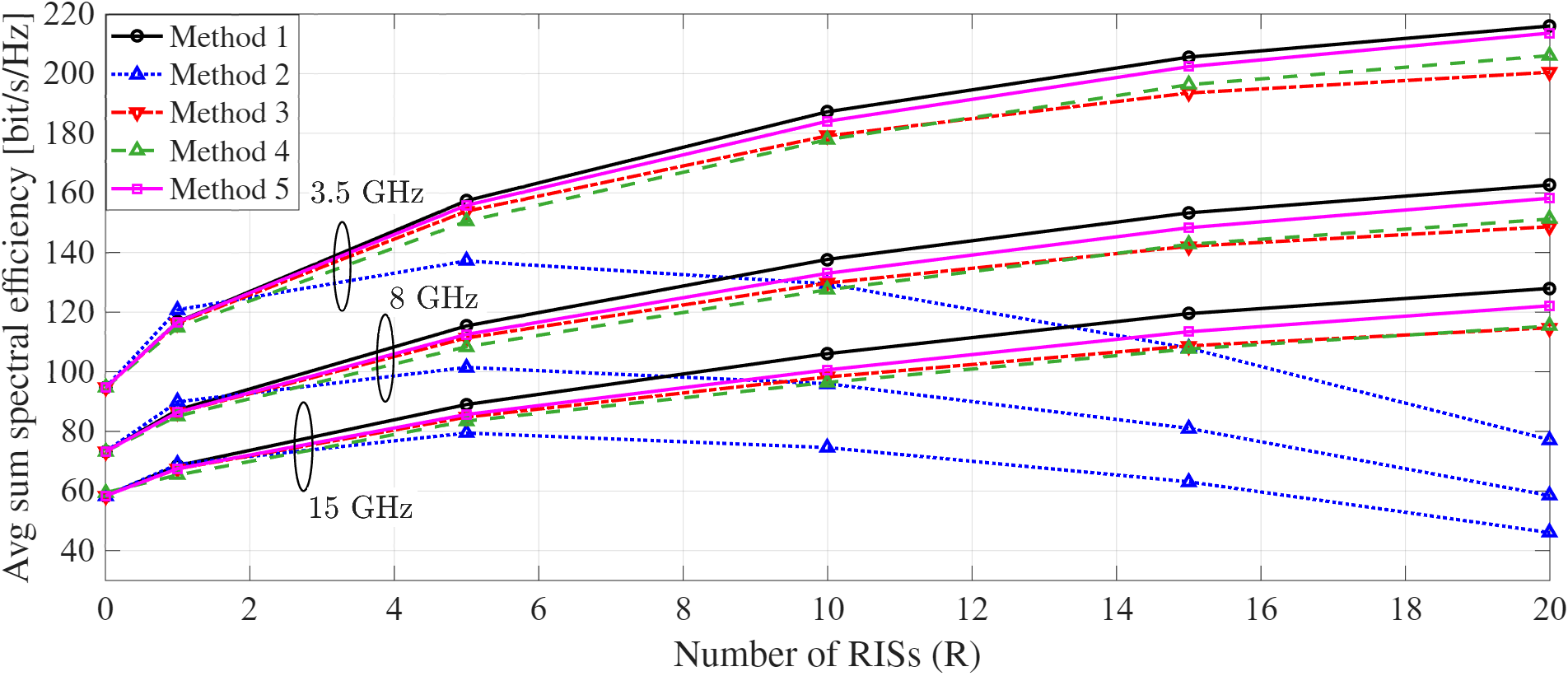}
\caption{Average sum spectral efficiency for $K=10$ MSs and $L=20$ APs as a function of the number of RISs ($R=0,\ldots,20$) under FR1 and FR3. The curves compare the five RIS-configuration methods described in Sec.~IV. 
\textcolor{black}{Method~5 implements 16-element subarray grouping, reducing the 
separation overhead.}}
\label{fig:SumSEvsRIS}
\end{figure}

\Cref{fig:SumSEvsRIS} reports the average sum \gls{se} for $K=10$ MSs as a function of the number of RISs $R\in\{0,\ldots,20\}$, for FR1 and FR3. 
We compare the following RIS configuration methods:
\begin{itemize}
    \item Method~1: LoS-aware RIS-MS assignment (one MS per RIS), with per-MS phase optimization. RISs without LoS to any MS remain unassigned and use a default low-overhead configuration.
    \item Method~2: Phase-shifts optimized using channels of all MSs.
    \item Method~3: Phase-shifts optimized using a randomly selected MS.
    \item \textcolor{black}{Method~4: Small pseudo-random codebook (no separation estimation).}
    \item \textcolor{black}{Method~5: LoS-aware RIS-MS assignment combined with \emph{subarray-based semi-blind sketching}. Each $64$-element RIS is partitioned into $N_\mathrm{B}=4$ blocks of $16$ elements, requiring only $1+N_\mathrm{B}=5$ separation states instead of $65$. This reduces $\tau_{\mathrm{p}_1}$ while preserving the dominant structure of the cascaded MS-RIS-AP path.}
\end{itemize}

\begin{figure*}[!t]
\subfloat[FR1: $3.5$ GHz, $K=10$ MSs]
{\includegraphics[width=\columnwidth]
{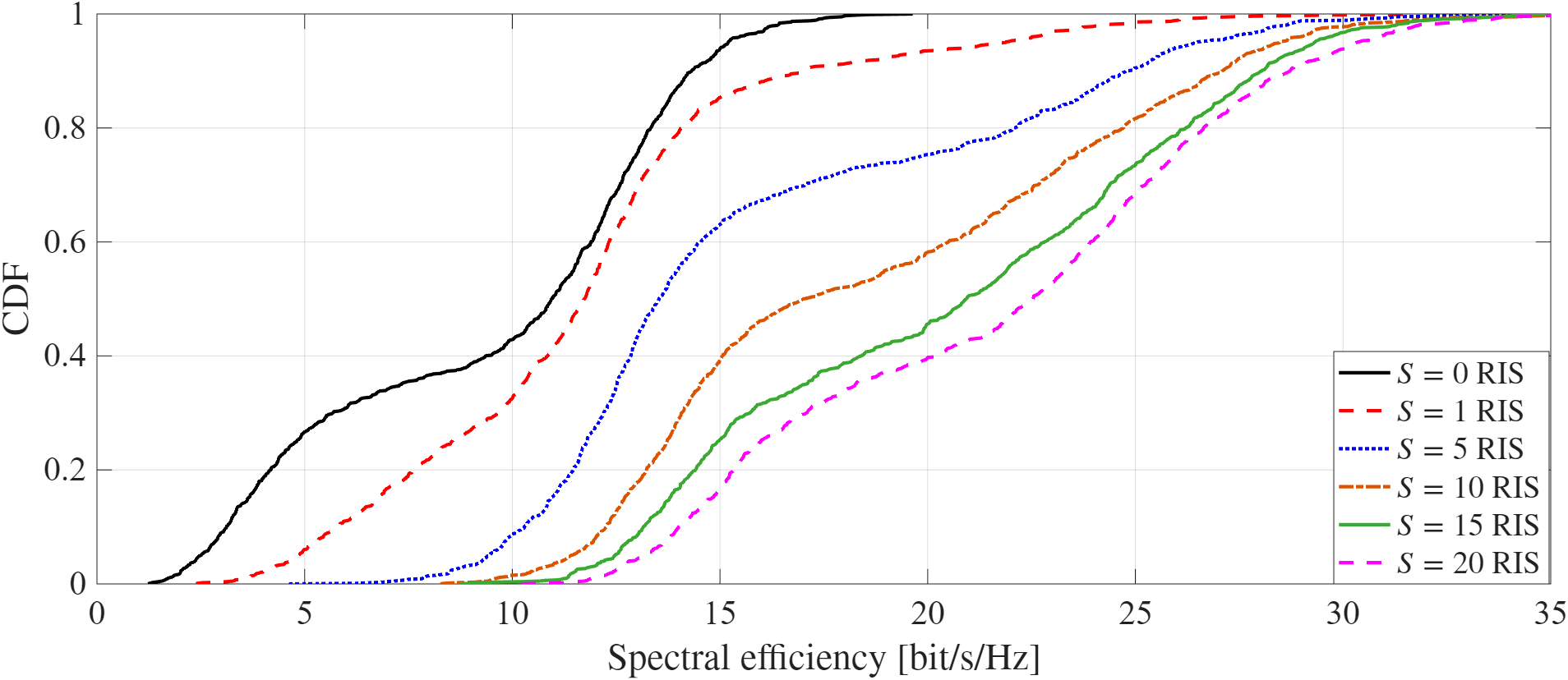}}\label{fig:3.5RIS}
\subfloat[FR1: $3.5$ GHz, $K=30$ MSs (stress test)]
{\includegraphics[width=\columnwidth]
{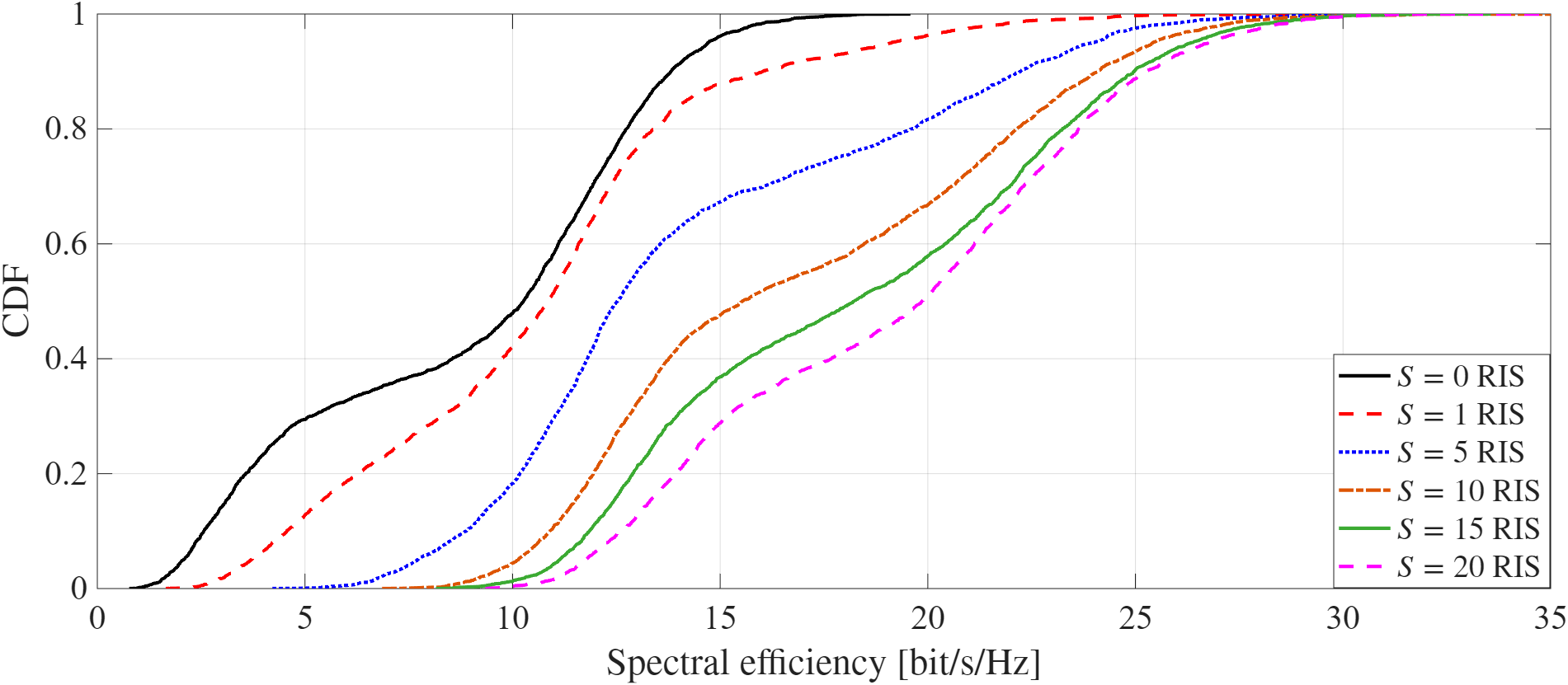}}\label{fig:30MS}\\
\subfloat[FR3: $8$ GHz, $K=10$ MSs]
{\includegraphics[width=\columnwidth]
{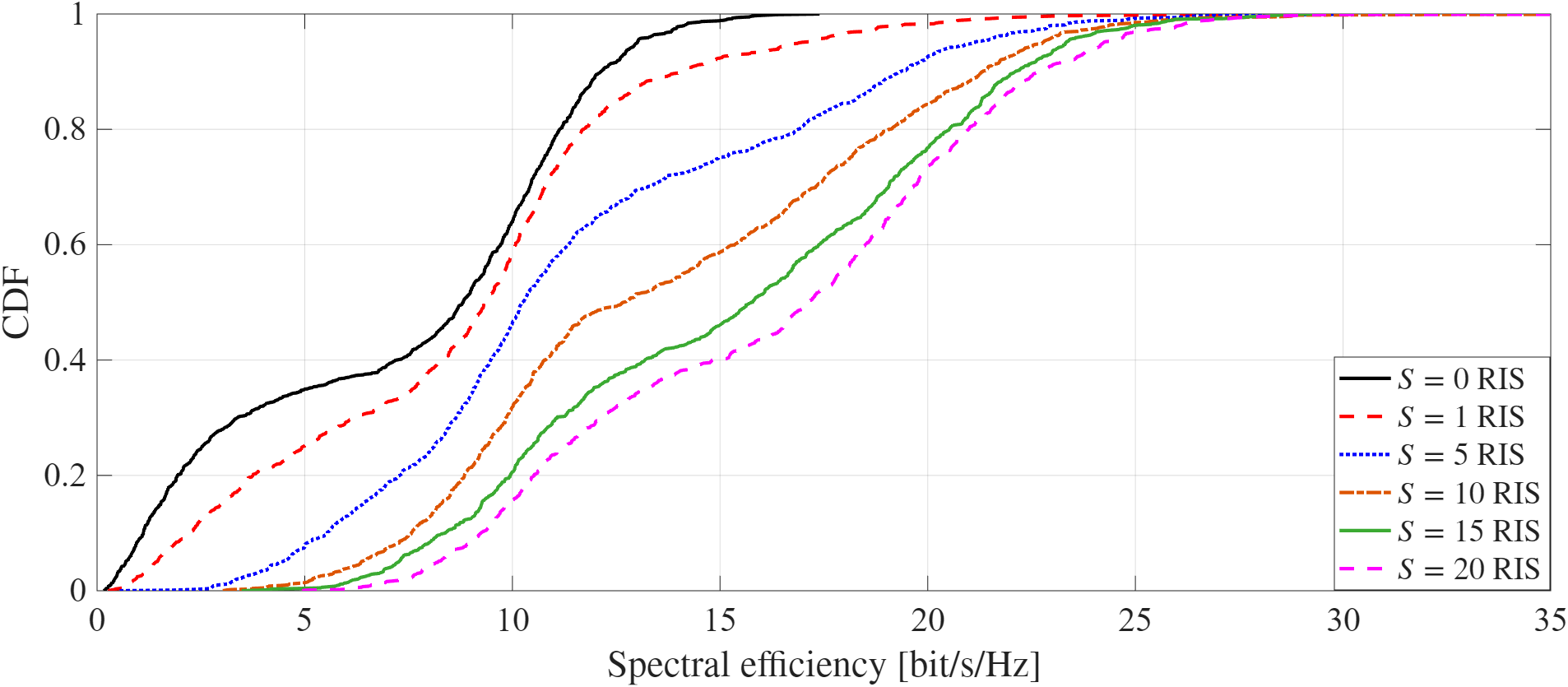}\label{fig:8RIS}} \
\subfloat[FR3: $15$ GHz, $K=10$ MSs]
{\includegraphics[width=\columnwidth]
{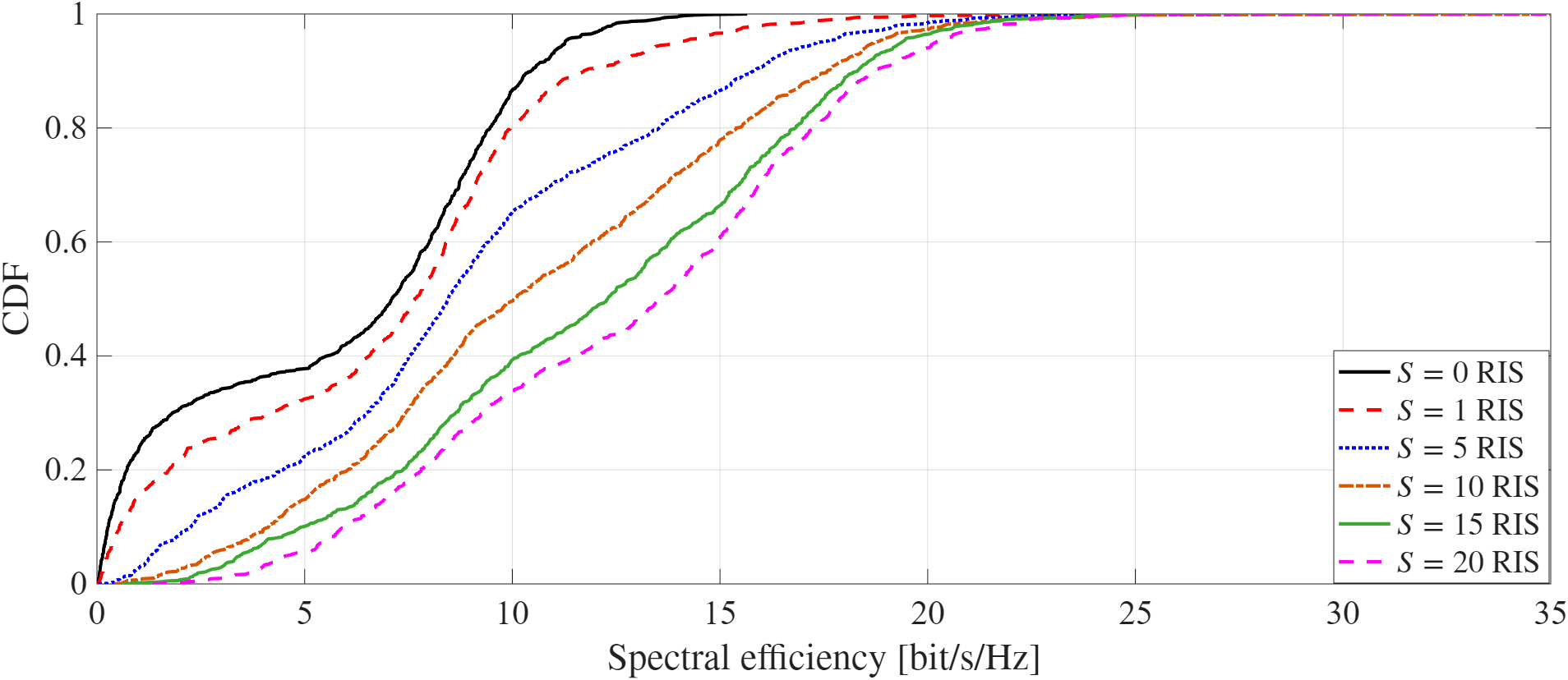}}\label{fig:15RIS}\\
\caption{CDF of the \gls{se} achieved with the proposed MS-RIS selection for $K=10$ MSs in a RIS-assisted CF-mMIMO network with P-MMSE combining. $L=20$ APs with $N_\mathrm{L}=4$ antennas, $R=\{0,1,5,10,15,20\}$ RISs with $N_\mathrm{R}=64$ metaatoms per RIS, and different frequency bands (FR1 and FR3). \textcolor{black} {High-density stress test with $K=30$ MSs illustrating the robustness of the LoS-aware RIS-MS selection under user crowding.}}
\label{fig:RIS_CDF}
\end{figure*}

Method~1 consistently outperforms Methods~3 and~4, and Method~2 suffers from 
excessive pilot overhead for large $R$. \textcolor{black}{Method~5 outperforms Method~1 when $R\geq5$ RISs independently of the frequency band, demonstrating that block-level grouping provides an excellent \gls{se}-overhead compromise.}

\textcolor{black}{
When additional RISs with LoS connectivity to the served MSs are introduced, new deterministic MS-RIS-AP components are created, reinforcing the direct MS-AP channel and substantially increasing the effective large-scale gain. These cascaded LoS-assisted paths also reduce the severity of small-scale fading and provide stronger spatial diversity to the combining stage. As a result, P-MMSE can suppress interference more effectively, leading to a pronounced improvement in both average and 
5th-percentile SE. This explains the significant performance jump observed when moving from $R=0$ to moderate values such as $R=5$. For larger $R$, the training overhead associated with RIS configuration begins to offset the channel-strengthening benefits, which is consistent with the saturation trend visible in Figure~\ref{fig:SumSEvsRIS}.}

\subsection{Per-MS SE distribution}

Figure \ref{fig:RIS_CDF} shows the \gls{cdf} of the per-MS \gls{se} for $K=10$ MSs and $R\in\{0,1,5,10,15,20\}$ RISs under FR1 and FR3 with $N_\mathrm{R}=64$ metaatoms. Increasing the number of RISs improves the per-MS \gls{se} across all percentiles, especially the 5th-percentile, since RISs with LoS links are directed to the weakest users. \textcolor{black}{The default configuration of unassigned RISs keeps the background channel stable while avoiding unnecessary separation overhead.}
Note that increasing from $R=15$ to $R=20$ offers little performance benefits since the channel estimation resource waste almost compensate the channel gain achieved by the introduction of $5$ additional \glspl{ris}. However, we should notice how with only one \gls{ris}, the \gls{se} per \gls{ms} in the system is significantly improved for the three frequency ranges, and the performance gain of increasing $R$ to $5,10,15$ \glspl{ris} is significant when an intelligent \gls{ris}-\gls{ms} selection method is employed.

\textcolor{black}{To evaluate the robustness of the proposed RIS-MS selection under high user density, Figure~2(b) reports a stress-test scenario with $K=30$ MSs. Despite the increased contention for RIS resources, the LoS-aware assignment continues to favor the weakest MSs, improving the 5th-percentile \gls{se} while preserving scalability, as each RIS is dedicated to a single MS and selection depends only on large-scale LoS connectivity.}

\subsection{Equal-area RIS panels and density scaling}

Despite the \gls{se} improvements observed employing a large number of \glspl{ris} independently of the frequency range, the poorer propagation environment using FR3 
derives in inferior \gls{se} performance when we employ \glspl{ris} with the same 
number of metaatoms. However, the element density scales with the square of the 
inverse of the wavelength, enabling a significantly higher number of elements to be 
accommodated within the same physical size as the operating frequency increases. 
To compare RIS panels of identical physical size ($30\times30$ cm), the number of 
reflective elements scales with frequency: $N_\mathrm{R}=64$ at $3.5$ GHz, 
$N_\mathrm{R}=256$ at $8$ GHz, and $N_\mathrm{R}=1024$ at $15$ GHz. 

\begin{figure}[!t]
\centering
\includegraphics[width=0.5\textwidth]{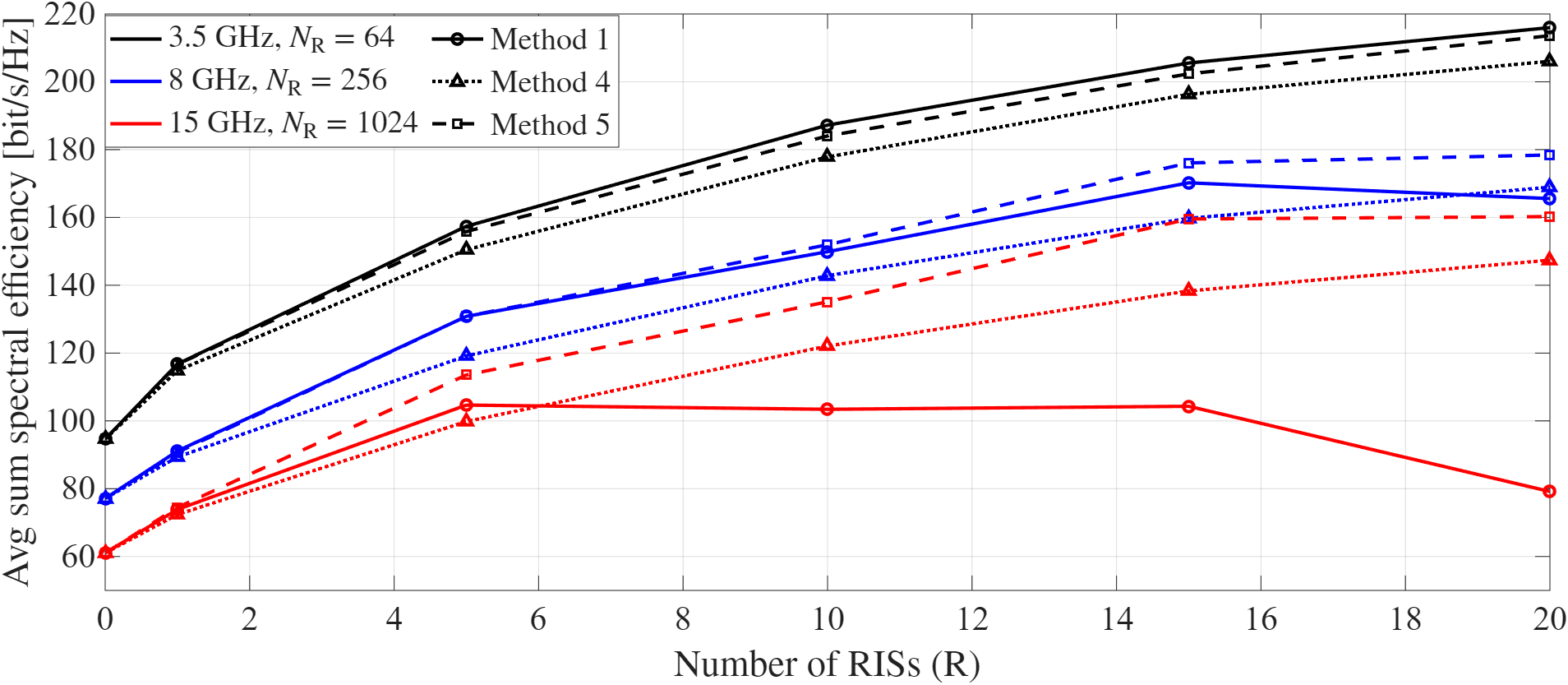}
\caption{Average sum spectral efficiency for equal-area RIS panels ($30\times30$ cm)
at $3.5$ GHz ($N_R = 64$), $8$ GHz ($N_R = 256$), and $15$ GHz \textcolor{black}{($N_R = 1024$). For each frequency, the figure reports the performance of Methods~1, 4, and~5 as a function of the number of RISs. The three curves per frequency illustrate the impact of conventional per-element configuration (Method~1), random low-overhead patterns (Method~4), and block-level grouping (Method~5) under identical physical RIS dimensions.}}
\label{fig:RISelements}
\end{figure}

\textcolor{black}{Figure~\ref{fig:RISelements} reports the sum SE versus $R$ for these three equal-area configurations, comparing Methods~1, 4, and~5 under identical physical RIS apertures.} As $N_\mathrm{R}$ increases with frequency, the potential RIS gain grows 
significantly, but so does the separation overhead, which may offset the benefits for 
large $R$ or dense arrays. It is worth noting that when the number of \glspl{ris} 
increases up to $R=15$ at $8$ GHz, the training overhead required to configure the 
\glspl{ris} with the proposed \gls{ris}-\gls{ms} selection Method~1 degrades the 
performance, and using randomly configured $R=20$ RISs with $N_\mathrm{R}=256$ metaatoms 
(Method~4) presents better sum \gls{se} results than configuring the \glspl{ris} with 
the proposed Method~1. This trend is exacerbated at higher frequencies; that is, at 
$15$ GHz, we employ $N_\mathrm{R}=1024$ metaatoms, which require more channel estimation 
resources, leading to a deterioration with a few \glspl{ris} that require estimating all 
the cascaded channels, even for a single user. 
\textcolor{black}{To address this limitation, Figure~\ref{fig:RISelements} includes 
\textbf{Method~5} for the three equal-area configurations ($N_\mathrm{R}=64,256,1024$). 
Each RIS is partitioned into blocks of $16$ elements, reducing the number of separation 
states from $1+N_\mathrm{R}$ to $1+N_\mathrm{R}/16$ (i.e., $5$, $17$, $57$). This alleviates the overhead that limits Method~1 when $N_\mathrm{R}$ grows, especially at FR3 frequencies where RISs can host hundreds of elements, and yields a substantially improved 
\gls{se}-overhead trade-off.}
These results indicate that dense RIS deployments in FR3 can yield substantial benefits, 
provided that overhead-aware configuration strategies are employed. \textcolor{black}{Subarray-based semi-blind sketching offers a scalable mechanism to operate large-aperture RISs efficiently within CF-mMIMO systems.}
 
\section{Conclusion}
\label{sec:conclusions}
This work confirms that RIS-assisted CF-mMIMO architectures can substantially enhance coverage and spectral efficiency in both FR1 and FR3 frequency bands. The proposed RIS-MS selection algorithm, based on LoS connectivity, effectively improves system performance while reducing channel estimation complexity compared to exhaustive approaches. Simulations reveal that RIS integration yields significant gains even with a small number of surfaces, and that increasing RIS density further enhances performance up to the point where pilot overhead becomes a limiting factor. Additionally, higher frequency bands (FR3) offer greater potential due to the ability to accommodate more RIS elements within the same physical dimensions, although this advantage depends on efficient channel estimation techniques. Future research should focus on advanced pilot compression and sparse estimation methods to fully exploit RIS capabilities in large-scale CF-mMIMO networks. Overall, intelligent RIS-user association and scalable processing are key enablers for realizing the benefits of RIS-assisted CF-mMIMO in next-generation wireless systems.

\bibliographystyle{IEEEtran}
\bibliography{IEEEabrv,main}
\end{document}